# FORECASTING DISEASE BURDEN IN PHILIPPINES: A SYMBOLIC REGRESSION ANALYSIS

Marvin G. Pizon
Agusan del Sur State College of Agriculture and Technology
mpizon@asscat.edu.ph

Emelyn F. Sagrado
Agusan del Sur State College of Agriculture and Technology
sagradoemelyn1@gmail.com

**Abstract**

The burden of disease measures the impact of living with illness and injury and dying prematurely, and it is increasing worldwide leading cause of death both global and national. This paper aimed to propose an index of diseases and evaluate a mathematical model to describe the burden of disease by cause in the Philippines from 1990 - 2016. Through Principal Component Analysis (PCA) the diseases categorized as: passed on diseases, vector born diseases, non-communicable diseases, accident, and intentional. Symbolic Regression Analysis was carried out. The study revealed that the number of disease burdens categorized using CPA would continue to decrease to the year 2020 except for non-communicable diseases.

## 1.0 Introduction

Reliable information on causes of death is crucial to the improvement of national and international health policies for the prevention and control of disease and injury. The study of Murray and Lopez (1997) revealed that reliable medical information is available for less than 30% of the estimated 50·5 million deaths that occur each year worldwide. However, other data sources can be used to develop cause-of-death estimates for populations. To be useful, assessments must be internally consistent, plausible, and reflect epidemiological characteristics suggested by community-level data.

Many empirical studies related on predicting the burden of disease, one of this is the study of Lopez and Mathers (2006) which projects the global burden of disease (GBD), it indicates that vascular diseases, HIV/AIDS, respiratory infections will remain leading causes of global disease burden and the projected increases in death and disability from tobacco use. Rapid population aging resulted in increases in some diseases' morbidity and prevalence rates, resulting in a series of significant scientific and social consequences, especially in medicine and healthcare (Koch, 2010).

In the Philippines settings, the high burden of disease from TB, large economic losses from mortality and morbidity from TB, and poor clinical outcomes all suggest that there is an urgent need for increased investment in TB control (Peabody et al., 2005). Buczak et al. (2014) developed a model to predict high and low dengue incidence to provide timely forewarnings in the Philippines. The total DALYs due to lung cancer in the Philippines were 38, 584 and the corresponding DALY rates per population of 1,000 were 0.4 and the burden of disease increased substantially (Morampudi et al., 2017).

The burden of diseases is an inevitable occurrence in human life. In 2004, the World Health Organization (WHO) had released a global estimated data, which is about 58.8 million cases. Several causes have been identified and explicitly ranked burden diseases; however, the data on its rate is changing annually. The most basic and critical thing we should do is to know more about the health condition to resolve these issues. Thus, studies were conducted to understand the causes of changes better and identify measures that possibly cause the increase or lower burden of disease rate.

## 2.0 Conceptual/Theoretical Framework

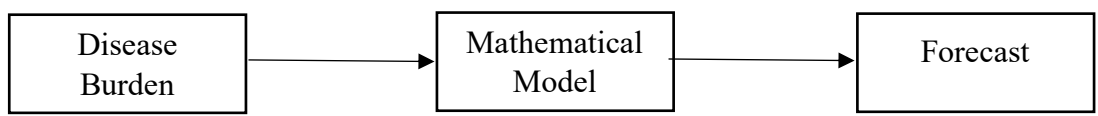

Figure 1. Conceptual Framework of the Study

The concept of the study was to forecast the burden of disease by cause in the Philippines in the future years that is presented in Figure 1. In this study, disease burden consists of three categories of disability or disease: non-communicable diseases (NCDs); communicable, maternal, neonatal, and nutritional diseases, and injuries. The symbolic regression model was utilized to determine the mathematical model with influential variables and estimate parameters in several stages, including operator selection, model solving, and model selection.

## 3.0 Research Design and Methods

In this study, total disease burden measured as the number of Disability-Adjusted Life Years (DALY's) per year was considered. DALYs are used to measure the total burden of disease by cause in which the number of years of the life of a person loses as a consequence of dying early because of the disease. One DALY equals one lost year of healthy life. And the data from 1990 - 2016 has been selected to study, the behavior of the disability or disease classification of diseases and injuries was based on Roser & Ritchie (2018).

In order to make the mathematical model valid, this paper investigates the multicollinearity among the independent variables because sometimes it is a problem that arises when there exists moderate to high intercorrelations among the predictor variables and information is redundant and can cause errors to be used in forecasting the data, and principal component analyses can cure this problem.

## 4.0 Results and Discussion

It is useful to look at the total proportion of variance in each eigenvector in deciding how many eigenvectors or principal components to be retained. As can be gleaned from table 4.1, three principal components were being considered and yielded a value of 63.44% and 18.1% on principal components 1 and 2, respectively, of the total variance. That is, almost 81% of the total variance is attributable to these two principal components. Moreover, figure 4.1 displays the scree plot on communicable diseases, and it appears that three principal component models were sufficient to represent the data set.

**Table 4.1 Principal Components based on Communicable Diseases.**

| Principal Component | 1 | 2 |
|---|---|---|
| | Coefficients | |
| HIV | -0.329 | 0.582 |
| Diarrhea | 0.451 | -0.122 |
| Malaria | 0.199 | 0.689 |
| Maternal | -0.222 | 0.248 |
| Neonatal | 0.424 | 0.320 |
| Nutritional | 0.451 | -0.022 |
| Other Communicable | 0.465 | 0.082 |
| Eigenvalue | 4.439 | 1.266 |
| Percentage of total variation explained | 63.4 | 18.1 |
| Cumulative percentage of total variation | 63.4 | 81.5 |

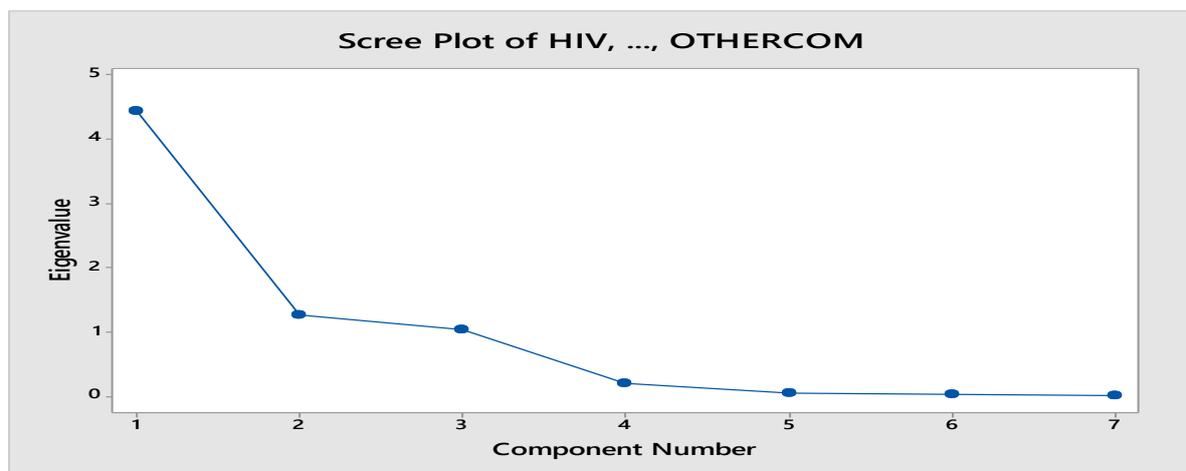

**Figure 4.1. Scree Plot on Communicable Disease**

Based on the results given in table 4.1, the first principal component is correlated to diarrhea, neonatal, nutritional, and other communicable diseases. This suggests that these four criteria vary together. If one increases, then the remaining one tends to as well. This component can be viewed as a measure of passed on disease (CPC1), and the second principal component is correlated to HIV, malaria, and maternal. And this component can be viewed as vector born disease (CPC2).

**Table 4.2 Principal Components based on Non-communicable Diseases.**

| Principal Component | 1 |
|---|---|
| | Coefficients |
| Cancers | 0.320 |
| Cardiovascular | 0.322 |
| Respiratory | 0.322 |
| Liver | 0.320 |
| Digestive | 0.283 |
| Neurology | 0.322 |
| Mental | 0.320 |
| Diabetes | 0.322 |
| Musculus | 0.318 |
| Other | 0.312 |
| Eigenvalue | 9.625 |
| Percentage of total variation explained | 96.2 |
| Cumulative percentage of total variation | 96.2 |

The total variance explained in the principal component, as presented in table 4.2, is clearly shown that only one principal component will be considered because having 96.2% will represent that this principle component was sufficient to represent the data set. And it can be seen from figure 4.2 displays the scree plot on communicable diseases, and it appears that only the principal component model was sufficient to represent the data set. Here we denote, principal component as a non-communicable disease (NPC) since all of the criteria are correlated.

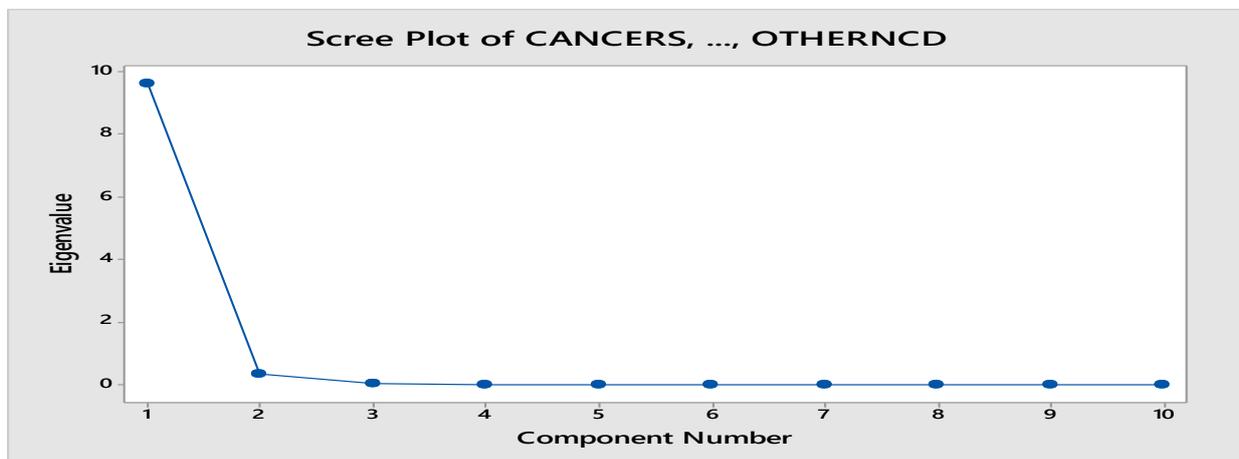

**Figure 4.2. Scree Plot on Non-communicable Disease**

As can be gleaned from Table 4.3, two principal components were being considered and yielded a value of 54.1% and 23.7% on principal components 1 and 2, respectively, of the total variance. That is, almost 77% of the total variance is attributable to these two principal components. Moreover, figure 4.1 displays the scree plot on injury cases, and it appears that two principal component models were sufficient to represent the data set. The first principal component for injury cases is correlated to three of the original variables. The principal component increases with increasing transport, interpersonal, and unintentional. This suggests that these three cases vary together and can be viewed as a measure of accident (IPC1), and for the second principal component of injury cases, natural, conflict, and self-harm vary together and can be viewed as a measure of intentional (ICP2).

**Table 4.3 Principal Components based on Injury Cases.**

| Principal Component | 1 | 2 |
|---|---|---|
| | Coefficients | |
| Transport | 0.532 | 0.168 |
| Natural | -0.154 | 0.620 |
| Conflict | -0.070 | 0.553 |
| Self-harm | -0.356 | 0.447 |
| Interpersonal | 0.542 | 0.132 |
| Unintentional | 0.517 | 0.256 |
| | | |
| Eigenvalue | 3.245 | 1.421 |
| Percentage of total variation explained | 54.1 | 23.7 |
| Cumulative percentage of total variation | 54.1 | 77.8 |

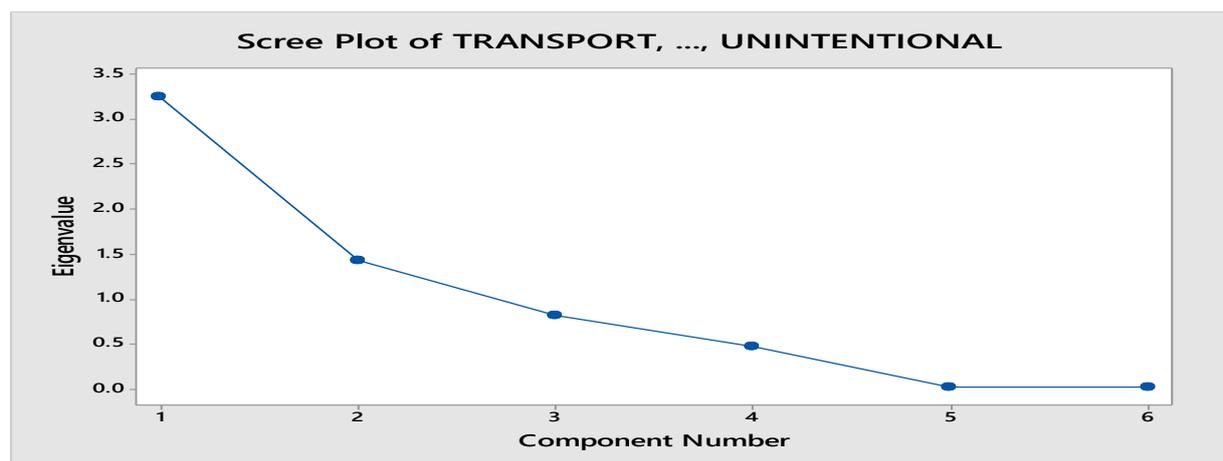

**Figure 4.3 Scree plot on Injury Cases**

The results of each index of the principal component will use to build a mathematical model that can be used to forecast the burden of disease, and symbolic regression was used to build a model—the summary of the solution of fit statistics of the model displayed in table 4.4. Model 1 is the mathematical equation that can be used to forecast passed on diseases, Model 2 for the vector born diseases, Model 3 for the non-communicable diseases, Model 4 for the accident, and Model 5 for the intentional. It can be gleaned from the table that all of the models have at least 0.99 of R2 indicate the percentage of forecast data that can be explained by the number of burden of diseases by cause. The R-value quantifies the correlation between the observed and the predicted values; all of the models are at least 0.99 of R, which is above 0.7 indicate the significant correlation between the model-estimated and actual output. The table further revealed that all of the models having at least 0.01 of Mean Absolute Error (MAE) and Mean Squared Error (MSE) estimates the difference between predicted and observed values and also MSE assumes that the error of measurement is normally distributed, since all of the model having at least 0.01 of

MAE and MSE and closer to zero, then it indicate the higher accuracy of predicting the values. Hence, all of the models are good in forecasting the number of burden diseases for the future year.

Figure 4.4 displays the plot of the actual data across the time (year), and the developed mathematical model was also shown. The graph of (a) – (e) illustrates the mathematical model 1 – 5, respectively. The effect of the graphs is evident to show that the data follows the developed mathematical model in each categorized burden of disease.

Table 4.4. Statistical Parameters for Mathematical Models

| Model | Mathematical Equations | $R^2$ | R | MSE | MAE | Complexity |
|---|---|---|---|---|---|---|
| 1 | $CPC1 = 6.31 + \dfrac{14.73}{t} + -\dfrac{14.59}{t^2} + 6.63\dfrac{\cos(t)}{t^3} - 1.63e - 6t^4$ | 0.9998 | 0.9999 | 0.0002 | 0.0115 | 37 |
| 2 | $CPC2 = 1.62 + 0.08t + 5.79e^{-5}t^3 + 0.04\cos(-12.96t) - 0.004t^2$ | 0.9973 | 0.9990 | 3.0003e$^{-5}$ | 0.0021 | 28 |
| 3 | $NPC = 9.32 + 0.16t + 0.02t^2 - 0.0005 * t^3$ | 0.9997 | 0.9998 | 0.0031 | 0.0411 | 19 |
| 4 | $IPC1 = 1.479 + 0.10t - 0.12\log(t) - 6.72e^{-5}t^3 - 0.0002t^2\sin(0.39t)$ | 0.9994 | 0.9997 | 6.23e$^{-5}$ | 0.0067 | 34 |
| 5 | $IPC2 = 0.78 + 0.006t^2 + 1.08e^{-7}t^5 - 0.10t - 6.59e^{-6} * t^4$ | 0.9999 | 0.9999 | 4.31e$^{-7}$ | 0.0006 | 33 |

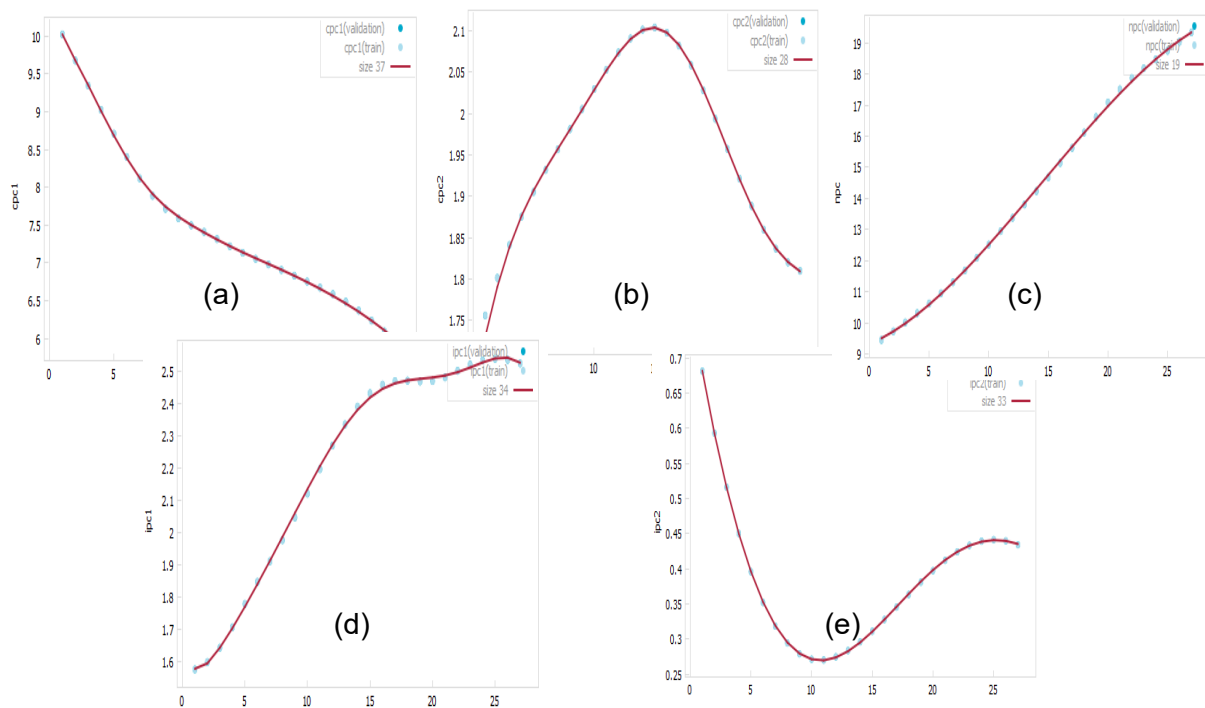

Figure 4.4. Graphs of the Mathematical Model

Given the mathematical model presented in table 4.4, the researcher will use the model to forecast the number of burden disease for the last four years, as illustrated in table 4.5. The data gathered were from 1990 – 2016, covering 17-year results by which the projection for the next five year-results are based. In table 4.5, the number burden of disease was calculated as a multiple of 103, and it shows that all of the disease burdens decreased up to the year 2020 except on the non-communicable disease. Non-communicable diseases (NCDs) are steadily increasing around the world, and developing countries are bearing much of occurring deaths. The study of Khalequzzaman et al. (2017) revealed a high prevalence of NCD was diabetes, dyslipidemia, hypertension, tobacco use, and both overweight and underweight were prevalent. Similar findings of Negi et al. (2016) that a high prevalence of overweight and obesity, hypertension, diabetes, and cardiovascular.

**Table 4.5. Computation results of forecasting on burden of disease (in millions) from 1990-2011**

|  | Year | CPC1 | CPC2 | NPC | IPC1 | IPC2 |
|---|---|---|---|---|---|---|
| Actual | 1990 | 10.03779 | 1.736238 | 9.453226 | 1.580009 | 0.655857 |
|  | 1991 | 9.642837 | 1.787833 | 9.698832 | 1.595836 | 0.868743 |
|  | . | . | . | . | . | . |
|  | . | . | . | . | . | . |
|  | . | . | . | . | . | . |
|  | 2015 | 6.083726 | 1.818974 | 19.03911 | 2.539031 | 0.330246 |
|  | 2016 | 5.946195 | 1.822348 | 19.34553 | 2.525819 | 0.421247 |
| Forecast | 2017 | 5.81208 | 1.8095 | 19.3422 | 2.52816 | 0.435552 |
|  | 2018 | 5.64442 | 1.80177 | 19.5646 | 2.48745 | 0.428418 |
|  | 2019 | 5.46157 | 1.79572 | 19.7494 | 2.41559 | 0.419002 |
|  | 2020 | 5.26203 | 1.78936 | 19.8938 | 2.30953 | 0.408052 |

**5.0 Conclusions**

In this paper, a symbolic regression approach is proposed and use to build a suitable model to forecast the disease burden in the Philippines. The proposed mathematical model has been analyzed and applied to the data of burden diseases for DALYs.

By the use of the principal component, the proposed model of burden diseases will categories as passed on diseases, vector born diseases, non-communicable diseases, accident and intentional and the developed mathematical model that is based on a statistical calculation of the parameters of mathematical models accomplish good results. It can be gleaned from the obtain results that all of the categorized burden diseases will be decreased up to the year 2020 except on the non-communicable diseases. Moreover, each corresponding model can be used as a basis for forecasting, planning, and management of the behavior of the number disease burden as the year increases.